\documentclass[prc,twocolumn,showpacs,preprintnumbers,amsmath,amssymb,superscriptaddress]{revtex4}

\usepackage{graphicx}
\usepackage{dcolumn}
\usepackage{bm}

\newcommand{\fo}{$^{18}$F}
\newcommand{\fn}{$^{19}$F}
\newcommand{\nen}{$^{19}$Ne}

\newcommand{\pa}{$^{18}$F(p,$\alpha)^{15}$O}
\newcommand{\dpt}{D($^{18}$F,p)$^{19}$F}
\newcommand{\dpa}{D($^{18}$F,p$\alpha)^{15}$N}
\def\power#1{\mbox{$\times10^{#1}\ $}}
\newcommand{\gap}{\mathrel{ \rlap{\raise.5ex\hbox{$>$}}
                    {\lower.5ex\hbox{$\sim$}}  } }
\newcommand{\lap}{\mathrel{ \rlap{\raise.5ex\hbox{$<$}}
		    {\lower.5ex\hbox{$\sim$}}  } }

\newcommand{\zaa}{{\it Astron. Astrophys.}}

\newcommand{\zapj}{{\it Astrophys. J.}}
\newcommand{\znp}{{\it Nucl.~Phys.}}

\newcommand{\zpr}{{\it Phys.~Rev.}}
\newcommand{\zprl}{{\it Phys.~Rev.~Lett.}}
\newcommand{\znim}{{\it Nucl.~Inst.~and~Meth.}}
\newcommand{\zADNDT}{{\it Atomic Data and Nuclear Data Tables}}
\newcommand{\zmnras}{{\it MNRAS}}
\newcommand{\zsitges}{{\it Proceedings of the International Conference on
		      Classical Nova Explosion}, Sitges, Spain, 20-24 May 2002,
		      		      AIP, 2002.}
				      
\begin{document}

\title{\dpa\ reaction applied to nova $\gamma$-ray emission}

\author{N.~de~S\'er\'eville}
\affiliation{CSNSM, IN2P3/CNRS and Universit\'e Paris-Sud, F-91405 Orsay
Campus, France}
\author{A.~Coc}
\affiliation{CSNSM, IN2P3/CNRS and Universit\'e Paris-Sud, F-91405 Orsay
Campus, France}
\author{C.~Angulo}
\affiliation{Universit\'e catholique de Louvain, Chemin du Cyclotron 2, 
B-1348 Louvain--la--Neuve, Belgium}
\author{M.~Assun\c{c}\~ao}
\affiliation{CSNSM, IN2P3/CNRS and Universit\'e Paris-Sud, F-91405 Orsay
Campus, France}
\author{D.~Beaumel}
\affiliation{IPN, IN2P3/CNRS and Universit\'e Paris-Sud, F-91406 Orsay Cedex,France}%
\author{B.~Bouzid}
\affiliation{USTHB, B.P. 32,  El-Alia, Bab Ezzouar, Algiers, Algeria}
\author{S.~Cherubini}
\thanks{Present address: EP3, Ruhr-Universit{\"a}t-Bochum, Bochum, Germany.}
\affiliation{Universit\'e catholique de Louvain, Chemin du Cyclotron
2, B-1348 Louvain--la--Neuve, Belgium}
\author{M.~Couder}
\affiliation{Universit\'e catholique de Louvain, Chemin du Cyclotron
2, B-1348 Louvain--la--Neuve, Belgium}
\author{P.~Demaret}
\affiliation{Universit\'e catholique de Louvain, Chemin du Cyclotron
2, B-1348 Louvain--la--Neuve, Belgium}
\author{F.~de~Oliveira~Santos}
\affiliation{GANIL, B.P. 5027, 14021 Caen Cedex, France}
\author{P.~Figuera}
\affiliation{Laboratori Nazionali del Sud, INFN, Via S. Sofia, 44 - 95123
Catania, Italy}
\author{S.~Fortier}
\affiliation{IPN, IN2P3/CNRS and Universit\'e Paris-Sud, F-91406 Orsay Cedex,France}%
\author{M.~Gaelens}
\affiliation{Universit\'e catholique de Louvain, Chemin du Cyclotron
2, B-1348 Louvain--la--Neuve, Belgium}
\author{F.~Hammache}
\thanks{Permanent address: IPN, IN2P3/CNRS and Universit\'e Paris-Sud,
F-91406 Orsay Cedex, France.}
\affiliation{GSI mbH, Planckstr. 1, D-64291 Darmstadt, Germany}
\author{J.~Kiener}
\affiliation{CSNSM, IN2P3/CNRS and Universit\'e Paris-Sud, F-91405 Orsay
Campus, France}
\author{A.~Lefebvre}
\affiliation{CSNSM, IN2P3/CNRS and Universit\'e Paris-Sud, F-91405 Orsay
Campus, France}
\author{D.~Labar}
\affiliation{Universit\'e catholique de Louvain, Chemin du Cyclotron 2,
B-1348 Louvain--la--Neuve, Belgium}
\author{P.~Leleux}
\affiliation{Universit\'e catholique de Louvain, Chemin du Cyclotron
2, B-1348 Louvain--la--Neuve, Belgium}
\author{M.~Loiselet}
\affiliation{Universit\'e catholique de Louvain, Chemin du Cyclotron
2, B-1348 Louvain--la--Neuve, Belgium}
\author{A.~Ninane}
\affiliation{Universit\'e catholique de Louvain, Chemin du Cyclotron
2, B-1348 Louvain--la--Neuve, Belgium}
\author{S.~Ouichaoui}
\affiliation{USTHB, B.P. 32,  El-Alia, Bab Ezzouar, Algiers, Algeria}
\author{G.~Ryckewaert}
\affiliation{Universit\'e catholique de Louvain, Chemin du Cyclotron
2, B-1348 Louvain--la--Neuve, Belgium}
\author{N.~Smirnova}
\affiliation{Instituut voor Kern en Stralingsfysika, Celestijnenlaan 200D,
B-3001, Leuven, Belgium}
\author{V.~Tatischeff}
\affiliation{CSNSM, IN2P3/CNRS and Universit\'e Paris-Sud, F-91405 Orsay
Campus, France}
\author{J.-P.~Thibaud}
\affiliation{CSNSM, IN2P3/CNRS and Universit\'e Paris-Sud, F-91405 Orsay
Campus, France}

\date{\today}

\begin{abstract}
The \pa\ reaction is recognized to be one of the most important reactions
for nova gamma--ray astronomy as it governs the early $E \leq$ 511~keV
gamma emission. However in the nova temperature regime, its rate remains 
largely uncertain due to unknown low--energy resonance strengths. We report 
here the measurement of the D($^{18}$F,p)$^{19}$F($\alpha)^{15}$N  
one--nucleon transfer reaction, induced by a 14~MeV \fo\  radioactive beam 
impinging on a CD$_2$ target; outgoing protons and $^{15}$N 
(or $\alpha$--particles) were detected in coincidence in two silicon strip 
detectors. A DWBA analysis of the data resulted in new limits to the 
contribution of low--energy resonances to the rate of the \pa\ reaction.
\end{abstract}

\pacs{25.60.Je, 21.10.Jx, 26.30.+k, 27.20.+n}
	
\maketitle

Gamma--ray emission from classical novae is dominated, during the first
hours, by positron annihilation resulting from the beta decay of radioactive 
nuclei. The main contribution comes from the decay of \fo\ (half--life of 
110~min) and hence is directly related to \fo\ formation during the 
outburst\cite{Gom98,Her99,CHJT00}. A good knowledge of the nuclear reaction 
rates of production and destruction of \fo\ is required to calculate the 
amount of \fo\ synthesized in novae and the resulting gamma--ray emission. The
rate relevant for the main mode of \fo\ destruction (i.e, through \pa) has 
been the object of many experiments\cite{Gra01,Bar02} (see also refs. in 
\cite{CHJT00}). However, this rate remains poorly known at novae 
temperatures (lower than 3.5\power{8}~K) due to the scarcity of spectroscopic 
information for levels near the proton threshold in the compound nucleus \nen.
This uncertainty is directly related to the unknown proton widths
($\Gamma_p$) of the first three levels ($E_x$, $J^\pi$ = 6.419~MeV, 3/2$^+$;
6.437~MeV, 1/2$^-$ and 6.449~MeV, 3/2$^+$). The tails of the corresponding 
resonances (at respectively $E_R$ = 8, 26 and 38~keV) can dominate the 
astrophysical S--factor in the relevant energy range \cite{CHJT00}. As a 
consequence of these nuclear uncertainties, the \fo\ production in novae and 
the early gamma--ray emission used to be uncertain by a factor of 
$\approx$300\cite{CHJT00}. Unfortunately, a direct measurement of the relevant
resonance strengths is impossible due to the very low Coulomb barrier 
penetrability. Hence, we used an indirect method aiming at the determination 
of the one nucleon spectroscopic factors ($S$) in the analog levels of the 
mirror nucleus (\fn) by a neutron transfer reaction: \dpt. Assuming the 
equality of one nucleon spectroscopic factors in analog levels, it is possible 
to calculate the proton widths through the relation 
$\Gamma_p=S\times\Gamma_{\mathrm s.p.}$ where
$\Gamma_{\mathrm s.p.}$ is the single particle width.
In this rapid communication we present the experimental technique, the
results of a DWBA analysis and the implications on the \pa\ thermonuclear
reaction rate.

The experiment has been carried out at the {\it Centre de Recherches du 
Cyclotron} at Louvain--la--Neuve (Belgium) where we used a \fo\ radioactive 
beam. The \fo\ was produced through the $^{18}$O(p,n)\fo\ reaction, 
chemically extracted to form CH$_3^{18}$F molecules, transferred to the 
cyclotron source, ionized to the $2^{+}$ state  and accelerated to 
14~MeV\cite{Cog99}. A total of 15 bunches of $\lap$1~Ci of \fo\ were produced 
providing each a mean beam intensity of 2.2$\times10^6$ particles per second 
over a period of $\approx$2~hours. The beam contamination from its stable 
isobar was very small, $^{18}$O / \fo\ $\le$ 10$^{-3}$. The \fo\ beam was 
directed on deuteriated polyethylene (CD$_2$) targets of $\approx$100~$\mu$g/cm$^2$ 
thickness made at Louvain--La--Neuve. For the energy considered here (1.4~MeV 
in the center of mass), the deuteron and the outgoing proton are both below 
the Coulomb barrier. In a similar transfer reaction $^{19}$F(d,p)$^{20}$F\cite{Lop64} 
studied at the same c.m. energy, strong direct contributions have been 
observed. Hence, with a stripping integral dominated by the best known portion
of the wave--function, accurate extraction of spectroscopic factors is
expected.

\begin{figure}[htb]
  \includegraphics[width=9cm]{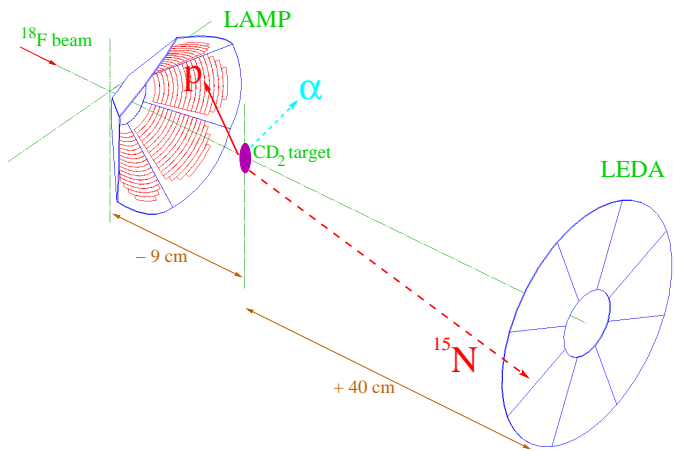}
  \caption{Experimental setup.}
  \label{f:setup}
\end{figure}

The experimental setup is depicted in Figure~\ref{f:setup}. It consists of 
two silicon multistrip detectors composed of sectors with 16 concentric 
strips\cite{Dav00}. The first detector, LAMP, is positioned 9~cm upstream 
from the target; it consists of 6 sectors forming a conical shape covering 
laboratory angles between 115$^\circ$ and 160$^\circ$ (i.e. forward 
center--of--mass angles between 12$^\circ$ and 40$^\circ$) providing a good 
acceptance for protons in the domain of interest for the differential cross 
section. The other detector, LEDA, is made up of 8 sectors forming a disk
positioned 40~cm downstream from the target and is used for background
reduction and normalization. The levels of interests are situated high above 
the alpha emission threshold (at 4.013~MeV) and their almost exclusive decay 
mode is through $^{19}$F$^*\to^{15}$N+$\alpha$. Hence, to reduce background, 
we required coincidences between a proton in LAMP and a $^{15}$N (or an 
$\alpha$--particle discriminated by time of flight) in LEDA. Following
Monte--Carlo simulations, the proton detection efficiency is found to be 27\% 
and is only slightly reduced to 19\% when the coincidence condition is applied.

\begin{figure*}[htb]
  \includegraphics[width=17cm]{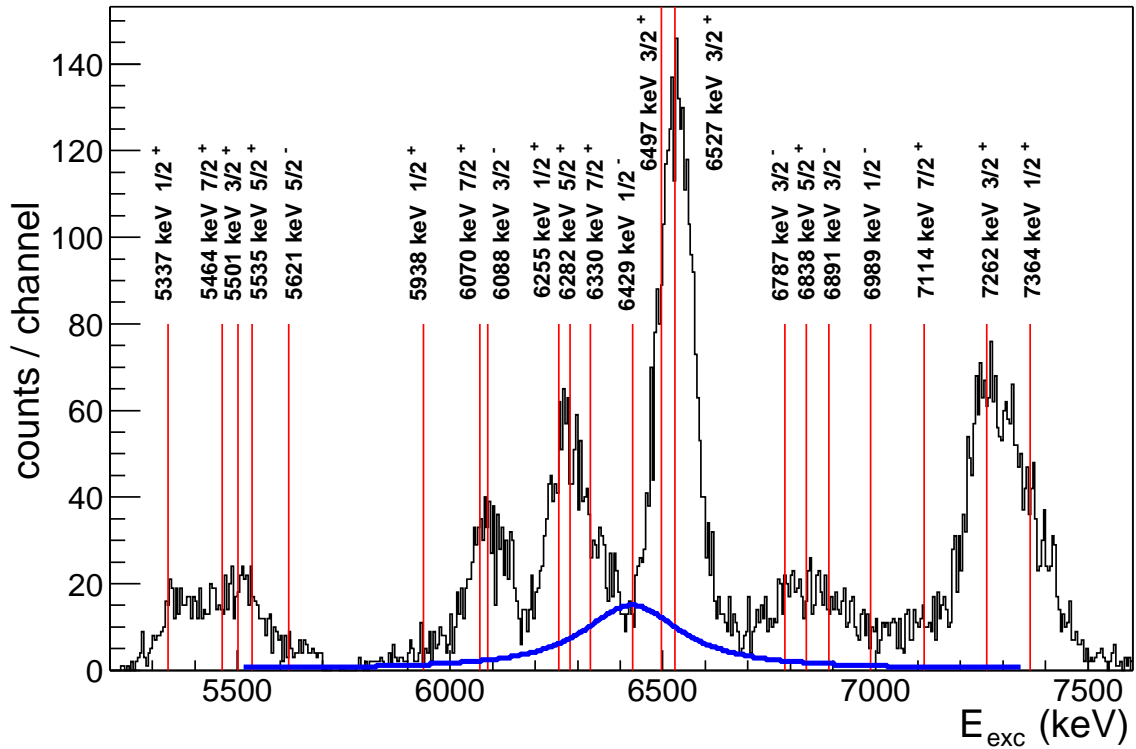}
  \caption{Example of a reconstructed \fn\ coincidence spectrum
  showing the two 3/2$^+$ levels of astrophysical interest around 6.5~MeV
  of excitation energy. The bold line shows the calculated contribution
  of the 1/2$^-$ level at 6429~keV for $S_3' = 0.15$ (see
  text). This spectrum is limited at low energy by the coincidence
  condition (no \fn\ breakup) and at high energy ($E_{X} \approx
  7.3$~MeV) by the proton energy below electronic threshold.}
  \label{f:spect}
\end{figure*}

Thanks to the kinematics, only protons and $\alpha$ from \dpt\ and 
D($^{18}$F,$\alpha)^{16}$O can reach LAMP. Using energy, angle and 
time--of--flight information, the events of interest with one proton in LAMP 
are selected. The excitation energy of the decaying \fn\ levels can be 
kinematically reconstructed from the energies and angles of the detected 
protons and the known beam energy. The energy calibration of the silicon 
strips has been done with a triple--line alpha source and a small correction 
($\approx$~1~\%) on the  excitation energy was introduced using a linear 
fit to the well--known energies of some lines in \dpt. This correction 
is needed because the reconstructed energy is very sensitive to the precise 
detector position (a 1~mm error induces a $\approx$~10~keV energy shift). The 
statistical error from the fitting procedure induced by this correction is 
about 5~keV at 6.5~MeV. The corresponding coincidence spectrum is shown in 
Figure~\ref{f:spect} where vertical lines correspond to the known position of 
the \fn\ levels\cite{Tilley} populated with low transferred angular momentum 
($l \leq 2$). The resolution ($\approx 100$ keV FWHM) is not sufficient to 
separate the various levels, but the two 3/2$^+$ levels of interest at 6.497 
and 6.528~MeV (the analogs of the 3/2$^+$ levels in \nen) are well separated 
from the other groups of levels.

\begin{figure}[htb]
  \includegraphics[width=9cm]{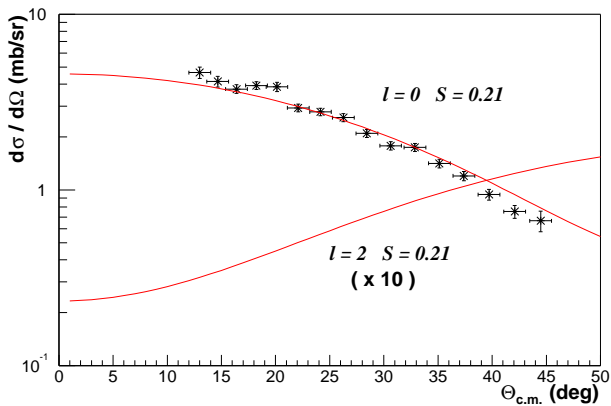}
  \caption{Comparison between the experimental angular distribution
  and DWBA calculations for different transferred angular momentum. The
  vertical error bars are only statistical whereas the horizontal ones
  are the angular width of each strip as seen from the target.}
  \label{f:dsig}
\end{figure}
   
The angular distribution shown in Figure~\ref{f:dsig} is the result of
a selection made on the 6.5~MeV peak in the coincidence spectrum. The 
coincidence efficiency is determined with a Monte-Carlo simulation taking
an isotropic angular distribution for the $\alpha$-decay of the \fn.
Coincidences were initially designed to eliminate
$^{18}$O(d,p)$^{19}$O events induced by a possible isobaric beam
contamination. This was not essential for this purpose because of the
high beam 
purity, but it was useful to reduce the background from electronic
noise affecting the low proton energy region. We checked that the ratio between
coincidence and single events was well reproduced by Monte--Carlo
simulations.
Rutherford elastic scattering of $^{18}$F on carbon from the target, detected
in LEDA, provides the (target thickness) $\times$ (beam intensity)
normalization. At such low beam intensity, the target stoichiometry is
not expected to be modified during the experiment. This was confirmed
by proton elastic scattering analysis of the targets done after the
experiment at the Orsay ARAMIS facility. The solid lines in Figure~\ref{f:dsig}
correspond to theoretical DWBA calculations made with the code 
FRESCO\cite{FRESCO} in the zero--range approximation for different transfered 
angular momentum ($l = 0, 2$). The nuclear potentials were taken from 
Ref.~\cite{Lop64} in which a similar neutron transfer reaction,
$^{19}$F(d,p)$^{20}$F, was studied at the same center--of--mass energy 
(subcoulomb transfer). A typical direct mechanism pattern was observed for 
11 angular distributions of $^{20}$F excited states. The comparison between 
our data and the calculations indicates a strong dominance of $l = 0$
transfer for the sum of the contribution of the two 3/2$^+$ levels and the
value obtained for the total spectroscopic factor is $S_{1}' + S_{2}'=0.21$
\footnote{Preliminary results from Kozub et al. quoted in a very recent 
paper\cite{Bar02} are in agreement with our value\cite{Sitges02}.}.
Using an alternate set of optical parameters from a
compilation\cite{Pey76} led to the same spectroscopic factor value
showing the stability of the result with respect to variation of the 
optical parameters.
Although no peak is seen in Figure~\ref{f:spect} for the 1/2$^-$ level 
($E_x$ = 6.429 MeV, $\Gamma$ = 280 keV) due to its large total width, it is 
possible to derive an upper limit for the spectroscopic factor of 
$S_{3}' \lap 0.15$ assuming an $l = 1$ transfer (see Figure~\ref{f:spect}).
The important result of this analysis is that the contribution of these
two $3/2^{+}$ resonances to the destruction rate of \fo\ {\em cannot} be
neglected.
Sources of uncertainty on spectroscopic factors are statistics
(negligible), normalization to elastic scattering data ($\approx$ 10~\%) but the major one is 
the DWBA method itself. This last one is difficult to evaluate but using
statistics of ratios between the directly measured proton widths
and the ones extracted from a DWBA analysis\cite{Ver90}, we estimate
this error to 25\% (r.m.s.) for spectroscopic factors greater than
0.01. Larger deviation occurs for smaller spectroscopic factor due to
the contribution of other mechanisms, but the high value of 0.21
provides a safe margin.

\begin{figure}[ht]
  \includegraphics[width=9cm]{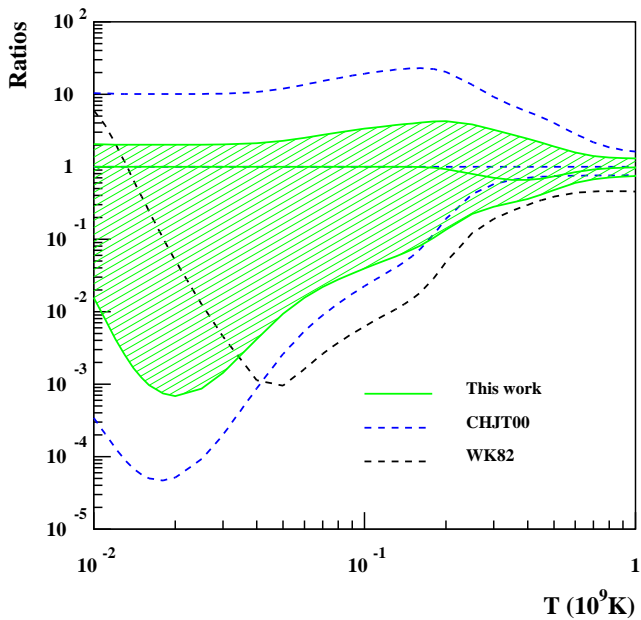}
  \caption{Present limits to the \pa\ reaction rate, calculated from
  this work and the new 330~keV resonance strength\protect\cite{Bar02} 
  (hatched area), compared to the previous
  situation\protect\cite{CHJT00} (CHJT00). All rates are
  normalized to the Coc et al. nominal rate\protect\cite{CHJT00} used
  in the more recent nova calculations. The Wiescher and Kettner
  rate\protect\cite{WK82} (WK82) used in the earliest nova 
  calculations\protect\cite{Gom98} is also displayed for comparison.}
  \label{f:rate}
\end{figure}

To show the improvement brought by this experiment, we present in
Figure~\ref{f:rate} the rates calculated in the same way as in
Ref.~\cite{CHJT00} except that we use new spectroscopic factors
obtained from this experiment  and the new value for the 330~keV
resonance strength\cite{Bar02}($\omega\gamma = 1.48 \pm 0.46$ eV).
Assuming the equality of spectroscopic factors between analog levels
one obtains $S_1+S_2$ = $S_1'+S_2'$ = 0.21 and $S_3$ = $S_3'<0.15$ where
$S_1'$, $S_2'$ and $S_3'$ correspond to the $E_x,J^{\pi}$ = 6.497~MeV,
$3/2^+$; 6.528~MeV, $3/2^+$; 6.429~MeV, $1/2^-$
levels in \fn\ while $S_1$, $S_2$ and $S_3$ correspond to the $E_x,J^{\pi}$ =
6.419~MeV, $3/2^+$; 6.449~MeV, $3/2^+$; 6.437~MeV, $1/2^-$ 
analog levels in \nen, respectively.
Taking into account the moderate uncertainty on the energy
calibration, the data favors a dominant contribution from the
6.528~MeV level i.e. $S_1' \approx 0.$ and $S_2' \approx 0.2$. However
the assignment of the two analog levels in \nen\ is not firmly
established. We cannot exclude inversion or mixing of these two $3/2^+$ 
levels. Hence to be conservative, for the calculation of the rate, we allow 
the individual spectroscopic factors to take any value compatible with
the $S_1+S_2$ = 0.21 and $S_3<0.15$ constraints. To show the
improvement provided by this experiment, we calculate
upper and lower rate limits by setting
$S_1=0$, $S_2=0.21$ (no inversion), $S_3=0.15$ and $S_1=0.21$, $S_2=0$
(inversion), $S_3=0$  respectively. Here we neglect the error associated with the spectroscopic factor
extraction discussed above.
As a consequence, the rate uncertainty is reduced by a factor of $\approx 5$
(Figure~\ref{f:rate}) in the temperature
range of novae. This is mainly due to the reduced contribution of the 1/2$^-$
level while the impact on the rate of the new contribution of the two 3/2$^+$
levels is more important at low temperatures.
It is worth noticing that these new rate limits still encompass the
nominal rate\cite{CHJT00} previously used in the most recent gamma--ray flux 
calculations. Since the repartition of spectroscopic factors between
the two $3/2^+$ levels is not settled, the choice $S_1=0.1$ and
$S_2=0.1$ for the nominal rate\cite{CHJT00} is still acceptable.
However, the new strength\cite{Bar02} of the resonance at 330~keV
has to be used for the rate calculation.

This work has been supported by the European Community-Access to Research
Infrastructure action of the Improving Human Potential Program, contract
N° HPRI-CT-1999-00110. One of us (P.L.) is a Research Director of the
National Fund for Scientific Research, Brussels.
We thank Alan Shotter and his team for allowing us to use the LEDA
and LAMP detectors from the Louvain-La-Neuve and Edinburgh collaboration.

\end{document}